\documentclass[superscriptaddress,nofootinbib,twocolumn,prl]{revtex4-1}

\usepackage{amsmath,amssymb}
\usepackage{graphicx}
\usepackage{color}
\usepackage{hyperref}
\usepackage{url}
\usepackage[utf8]{inputenc}
\usepackage{slashed}
\usepackage{subfigure}
\usepackage{slashed,bbm}
\usepackage{graphics,psfrag,epsfig}
\usepackage{dsfont}
\usepackage{setspace}

\def\code#1{\texttt{#1}}

\begin{document}

\title{Gross-Neveu-Yukawa model at three loops and Ising critical behavior of Dirac systems}

\author{Luminita~N.~Mihaila}
\affiliation{Institut f\"ur Theoretische Physik, Universit\"at Heidelberg, Philosophenweg 16, 69120 Heidelberg, Germany}

\author{Nikolai~Zerf}
\affiliation{Institut f\"ur Theoretische Physik, Universit\"at Heidelberg, Philosophenweg 16, 69120 Heidelberg, Germany}

\author{Bernhard~Ihrig}
\affiliation{Institut f\"ur Theoretische Physik, Universit\"at Heidelberg, Philosophenweg 16, 69120 Heidelberg, Germany}

\author{Igor~F.~Herbut}
\affiliation{Department of Physics, Simon Fraser University, Burnaby, British Columbia, Canada V5A 1S6}

\author{Michael~M.~Scherer}
\affiliation{Institut f\"ur Theoretische Physik, Universit\"at Heidelberg, Philosophenweg 16, 69120 Heidelberg, Germany}

\begin{abstract}
Dirac and Weyl fermions appear as quasi-particle excitations in many different condensed-matter systems.
They display various quantum transitions which represent unconventional universality classes related to the variants of the Gross-Neveu model.
In this work we study the bosonized version of the standard Gross-Neveu model -- the Gross-Neveu-Yukawa theory -- at three-loop order, and compute critical exponents in $4-\epsilon$ dimensions for general number of fermion flavors. Our results fully encompass the previously known two-loop calculations, and agree with the known three-loop results in the purely bosonic limit of the theory.
We also find the exponents to satisfy the emergent super-scaling relations in the limit of a single-component fermion, order by order up to three loops. 
Finally, we apply the computed series for the exponents and their Pad\'e approximants to several phase transitions of current interest: metal-insulator transitions of spin-1/2 and spinless fermions on the honeycomb lattice, emergent supersymmetric surface field theory in topological phases, as well as the disorder-induced quantum transition in Weyl semimetals. Comparison with the results of other analytical and numerical methods is discussed. 
\end{abstract}

\maketitle

\paragraph{Introduction.}


Dirac and Weyl fermions are an abundant form of quasi-particle excitations in condensed matter physics\cite{wehling2014,2013arXiv1306.2272V} appearing in very different materials ranging from graphene, via $d$-wave superconductors, to the surface states of topological insulators, or even three-dimensional (3D) materials such as Na${}_3$Bi and Cd${}_3$As${}_2$\cite{Liu864,PhysRevLett.113.027603}.
While the physical origin of the quasi-relativistic energy dispersion can be quite different in various materials, it leads to universal low-energy properties shared by all these materials, such as, e.g., the density of states (DOS) and the concomitant thermodynamic properties or various response functions.
Dirac and Weyl systems can also undergo transitions from their natural semi-metallic phase to a variety of broken symmetry phases as some parameter is varied.
This includes continuous quantum transitions to interaction-induced, ordered, many-body ground states\cite{sorella1992,Khveshchenko2001,herbut2006,honerkamp2008,herbut2009,raghu2008} and a disorder-driven transition to a diffusive state with finite DOS\cite{PhysRevB.33.3263,PhysRevB.33.3257,PhysRevLett.107.196803,PhysRevLett.112.016402,PhysRevB.94.220201,PhysRevB.93.155113}.
Depending on the broken symmetry of the ordered phase and the fermionic content, the corresponding transition represents a universality class with the concomitant critical behavior.
Various of these transitions have been suggested to belong to the universality class defined by the chiral transition appearing in the 3D Gross-Neveu (GN) model\cite{herbut2006} well-known in the context of high-energy physics and conformal field theories\cite{PhysRevD.10.3235, doi:10.1093/ptep/ptw120}.
This applies, in particular, to the interaction-induced transition toward a charge density wave of electrons on the 2D honeycomb lattice that breaks the (Ising) sublattice symmetry\cite{herbut2006}, or the disorder-driven transition toward a diffusive metal in a 3D Weyl semi-metal\cite{PhysRevB.94.220201}.
%

\paragraph{Model.}

The bosonized GN model -- the Gross-Neveu-Yukawa (GNY) model -- is represented by the Lagrangian
\begin{align}\label{eq:lagrangian}
	\mathcal{L}=\bar\psi(\slashed{\partial}+g\phi)\psi+\frac{1}{2}\phi(m^2-\partial_\mu^2)\phi+\lambda\phi^4\,,
\end{align}
which is defined in Euclidean space and is renormalizable in $D=4-\epsilon$ dimensions. It includes a real scalar field $\phi$ resulting from a Hubbard-Stratonovich decoupling of the four-Fermi interaction and lies in the same universality class as the GN model for (space-time) dimensions $2 < D < 4$\cite{HASENFRATZ199179,ZINNJUSTIN1991105}.
Here, $\slashed{\partial}=\gamma_\mu\partial_\mu$ and we use a four-dimensional representation of the Clifford algebra $\{\gamma_\mu,\gamma_\nu\}=2\delta_{\mu\nu}\mathbbm{1}_4$, with $\mu, \nu, = 0,1,...D-1$. 
The conjugate of the Dirac field is given by $\bar\psi=\psi^\dagger\gamma_0$.
The scalar field couples to the fermions with the Yukawa coupling $g$ and has a self-interaction (quartic term coupling) $\lambda$. For generality, we allow for an arbitrary number $N$ of four-component fermion species.

The precise determination of the critical exponents of the GN model is a formidable task and has been attempted by various methods, e.g., perturbative\cite{Rosenstein:1993zf,Zerf:2016fti,GRACEY1990403,Gracey:2016mio} and non-perturbative\cite{PhysRevB.94.245102,PhysRevB.89.205403} renormalization group (RG) approaches, Monte Carlo (MC) simulations\cite{KARKKAINEN1994781,PhysRevD.88.021701,1367-2630-16-10-103008,1367-2630-17-8-085003,PhysRevD.88.021701,PhysRevB.93.155157} and the conformal bootstrap, see Refs.~\cite{Poland:2016chs,Bashkirov:2013vya,Iliesiu:2015akf}.
For the case of the purely bosonic~$\phi^4$ theory with Ising symmetry in 3D, the development and comparison of these different methods has led to an impressive convergence across different theoretical approaches, settling beyond a three digit agreement, for example, for the correlation length exponent $\nu_\mathrm{Ising}\approx 0.630$\cite{0305-4470-31-40-006,PhysRevE.65.066127,PhysRevE.68.036125,PhysRevB.82.174433,Kos2016}.
For the 3D Gross-Neveu, or ``chiral Ising" universality class, the calculation of critical exponents has also been attempted by different methods; here, however, the situation is less settled, cf. Tab.~\ref{tab:critexp}.
Recent progress in MC simulations\cite{PhysRevD.88.021701,1367-2630-16-10-103008,1367-2630-17-8-085003,PhysRevD.88.021701} and the application of field-theoretical methods\cite{Gracey:2016mio,PhysRevB.94.245102,PhysRevB.89.205403} could not resolve the existing discrepancies, and the difference between the results still shows up in the first digits.
In fact, the paradigmatic role of the GNY model notwithstanding, its critical exponents are currently only known to two-loop order in the epsilon expansion near four dimensions\cite{Rosenstein:1993zf}.

We have calculated the beta-functions and the critical exponents for the GNY model with the general number of fermion flavors~$N$ at three-loop order, providing an important further step towards a more quantitative understanding of the fermionic universality classes.
The full analytical expressions for general $N$ are presented below.
Numerical values for the physically relevant cases $N=2,1,1/4$ and $0$ are presented in Tab.~\ref{tab:critexp2}, where we display the first three terms in the expansion in $\epsilon$ for the correlation length exponent and the anomalous dimensions.
Furthermore, we use Pad\'e approximants to extract estimates for the critical exponents at $\epsilon=1$.

The rest of the paper is organized as follows. First, we specify the RG procedure, the employed computer algebraical tools, and present the full set of three-loop RG functions. We then determine the fixed point solutions to order $\epsilon^3$ and calculate the anomalous dimensions and the correlation length exponent to that order. For the specific cases of physical interest $N=2,1,1/4,0$ we also calculate the Pad\'e approximants for the universal critical exponents and discuss the physical applications for quantum phase transitions in Dirac and Weyl semimetals. Finally, we draw our conclusions.

\paragraph{RG procedure and tools.}

We first explain how we set up the three-loop RG analysis in $D=4-\epsilon$ space-time dimensions.
The bare Lagrangian is defined by Eq.~\eqref{eq:lagrangian} upon replacing the fields and couplings by their bare counterparts $\psi\to \psi_0, \phi\to \phi_0, g\to g_0, \lambda \to \lambda_0$. The renormalized Lagrangian is then introduced as
\begin{align}
	\mathcal{L}=&Z_\psi\bar\psi\slashed{\partial}\psi-\frac{1}{2}Z_\phi(\partial_\mu\phi)^2+Z_{\phi^2}\frac{m^2}{2}\phi^2\nonumber\\
	&+Z_{\phi\bar\psi\psi} g \mu^{\epsilon/2}\phi\bar\psi\psi+Z_{\phi^4}\lambda\mu^\epsilon\phi^4\,,
\end{align}
where $\mu$ defines the energy scale which parametrizes the RG flow of the couplings. We have defined the wave function renormalization constants $Z_\psi$ and $Z_\phi$ which relate the bare and the renormalized Lagranigan upon the field recaling $\psi_0=\sqrt{Z_\psi}\psi$ and $\phi_0=\sqrt{Z_\phi}\phi$.
The explicit $\mu$ dependencies in $\mathcal{L}$ reflect that after introducing the integration over $D=4-\epsilon$ dimensional spacetime we shift $g^2\,\to g^2\mu^\epsilon$ and $\lambda\to\lambda\,\mu^\epsilon$. To simplify the notation in the following we introduce $y=g^2$. The renormalization constants for the mass term, the Yukawa coupling and the quartic coupling are further introduced by 
\begin{align}\label{eq:Z}
	m^2&=m_0^2Z_\phi Z_{\phi^2}^{-1}\,,\\
	y&=y_0\mu^{-\epsilon}Z_\psi^2Z_\phi Z_{\phi\bar\psi\psi}^{-2}\,,\quad \lambda=\lambda_0\mu^{-\epsilon}Z_\phi^2Z_{\phi^4}^{-1}\label{eq:Z2}\,.
\end{align}
These relations provide the RG scale dependence of the renormalized quantities by taking into account the RG invariance of the bare quantities.
We calculate the renormalization group constants $Z_x$ where $x \in\{\psi,\phi,\phi^2,\phi\bar\psi\psi,\phi^4\}$ up to three-loop order employing dimensional regularization and the modified minimal subtraction scheme ($\overline{\text{MS}}$).
To that end, we use sophisticated computer algebra which was established for higher-loop calculations in high-energy physics, in particular in the context of Standard Model of Particle Physics calculations: The complete set of Feynman diagrams is generated with the program \code{QGRAF}\cite{NOGUEIRA1993279} and further processed with the programs \code{q2e} and \code{exp}\cite{Harlander1998125,Seidensticker:1999bb}.
Traces over matrix structures coming from the Clifford algebra and the tensor reduction of Feynman intergrals is then achieved with \code{FORM}\cite{Vermaseren:2000nd,Kuipers20131453}.
The calculation of Feynman integrals is performed after reduction to master integrals via integration-by-parts identities.
We evaluate the beta functions and the anomalous dimensions using two independent setups.
In one of them we computed the vertex functions setting one or two external momenta to zero.
In this case, the loop integrals are mapped to massless two-point functions that are implemented up to three loops in the code \code{MINCER}\cite{Larin:1991fz}.
In the second setup we introduce an infrared regulator for all the propagators as described in Ref.~\cite{Chetyrkin:1997fm}.
Here, the loop integrals can be reduced to three-loop tadpole integrals that can be processed with the help of \code{MATAD}\cite{STEINHAUSER2001335}.

\paragraph{Beta functions.}

The beta functions for the squared Yukawa coupling $y$ and the quartic scalar coupling $\lambda$ are defined as $\beta_y=dy/d\ln \mu$ and $\beta_\lambda=d\lambda/d\ln \mu$. The relation to the renormalization constants is derived from Eqs.~\eqref{eq:Z}-\eqref{eq:Z2}. We work with rescaled couplings $y/(8\pi^2) \to y$ and $\lambda/(8\pi^2)\to \lambda$.
The beta functions for the Yukawa and the quartic scalar coupling at three-loop order read
\begin{align}
	\beta_{y}=&-\epsilon y+(3 +2 N)y^2+24y\lambda(\lambda- y)-\big(\frac{9}{8}+6N\big)y^3\nonumber\\
	&+\frac{y}{64}\Big(1152 (7+5N) y^2\lambda+192 (91-30 N) y \lambda ^2\nonumber\\
	&+\big(912 \zeta_3-697+2 N (67+112 N+432 \zeta_3)\big)y^3\nonumber\\
	&-13824 \lambda ^3\Big)\,,\\[3pt]
	\beta_{\lambda}=&-\epsilon  \lambda +36 \lambda ^2+4N y \lambda -N y^2+4 N y^3+7N y^2 \lambda\nonumber\\
	& -72N y \lambda ^2-816 \lambda ^3+\frac{1}{32} \Big(6912 (145+96 \zeta_3) \lambda ^4\nonumber\\
	&+49536 N y \lambda ^3-48 N (72 N-361-648 \zeta_3)y^2\lambda ^2\nonumber\\
	&+2 N (1736 N-4395-1872 \zeta_3)y^3 \lambda\nonumber\\
	&+N (5-628 N-384 \zeta_3)y^4\Big)\,,
\end{align}
where $\zeta_z$ is the Riemann zeta function.
Our expressions fully agree up to two loops with the ones from\cite{Rosenstein:1993zf}. Upon setting $y=0$, the beta function for the quartic coupling also agrees with the three-loop results for the real scalar $\phi^4$ theory with $\mathbbm{Z}_2$ or Ising symmetry\cite{Kleinert:2001ax}.

\paragraph{Fixed points.}

The three-loop beta functions allow the determination of the RG fixed points of the system order by order in $\epsilon$ up to order $\epsilon^3$.
 Let us start the fixed-point analysis at the one-loop level, where the beta functions for $y$ and $\lambda$ give rise to four different fixed points\cite{PhysRevB.80.075432}: the unstable Gau\ss ian fixed point with vanishing coordinates $(y_\ast,\lambda_\ast)_0=(0,0)$, the unstable bosonic Wilson-Fisher fixed point $(y_\ast,\lambda_\ast)_\text{WF}=(0,\epsilon/36)$, and a pair of fully non-Gau\ss ian fixed points~(NGFP)
\begin{align}\label{eq:NGFP}
(y_\ast,\lambda_\ast)_{\pm}=\left(\frac{1}{3+2N}\,\epsilon,\frac{3-2N\pm s}{72(3+2N)}\,\epsilon\right)\,,
\end{align}
where $s=\sqrt{9+4N(33+N)}$. From the pair of NGFPs, the one with the negative solution is discarded as it has a negative quartic coupling.
Here, we study the stable positive solution from Eq.~\eqref{eq:NGFP}, which we solve order by order in $\epsilon$.
The expressions for general $N$ are lengthy at order $\epsilon^3$ and we refrain from fully displaying them here.
Instead, we explicitly show the universal critical exponents at order $\epsilon^3$ as derived from the fixed-point solution.

\paragraph{Critical exponents.}

The field renormalization constants are defined as the logarithmic derivatives of the wave function renormalizations of the fermion and the boson fields, $\gamma_x=d\ln Z_x/d\ln \mu$ for $x\in \{\psi,\phi\}$ and read
\begin{align}
	\gamma_\psi=&\frac{y}{2}-\frac{1}{16}(1+12N)y^2+\frac{y}{128}\Big((48 \zeta_3-15\nonumber\\
	&+4N(47-12N))y^2+768 y \lambda -2112  \lambda ^2\Big)\,,\\
	\gamma_\phi=&2Ny+24 \lambda ^2-\frac{5 N y^2}{2}+\frac{1}{32}\Big(N y^3 (21+200 N\nonumber\\
	&+48 \zeta_3)+960 N y^2 \lambda -2880 N y \lambda ^2-6912 \lambda ^3\Big)\,.
\end{align}
Our expressions agree with the ones from\cite{Rosenstein:1993zf} and \cite{Kleinert:2001ax} in the corresponding limits.
To obtain the fermion and boson anomalous dimensions characterizing the critical behavior, we evaluate these expressions at the NGFP and define $\eta_\psi=\gamma_\psi(y_\ast,\lambda_\ast),\ \eta_\phi=\gamma_\phi(y_\ast,\lambda_\ast)$, see below.

Finally, we require yet another renormalization constant, i.e. the one related to the renormalized mass term.  We define $\gamma_{\phi^2}=d \ln Z_{\phi^2}/d\ln\mu$. At three-loop order,
\begin{align}
&\gamma_{\phi^2}=-2(6\lambda +N y^2-12 N y \lambda -72 \lambda^2)-72 \lambda ^2 (4N y+87 \lambda )\nonumber\\
&-4N y^3 (4 N-9+3 \zeta_3)-\frac{3}{2}N y^2 \lambda(11-24N+120\zeta_3).\nonumber
\end{align}
The RG beta function of the dimensionless mass term $\tilde m^2=\mu^{-2}m^2$, then follows from Eq.~\eqref{eq:Z} and reads $\beta_{\tilde m^2}=(-2+\gamma_\phi-\gamma_{\phi^2})\tilde m^2$. We extract the correlation length exponent at the stable NGFP $(y^\ast,\lambda^\ast)$ from the relation
\begin{align}
\nu^{-1}= \theta_1=-\frac{d\beta_{\tilde m^2}}{d \tilde m^2}\Big|_{(y^\ast,\lambda^\ast)}=2-\eta_\phi+\eta_{\phi^2}\,,
\end{align}
where $\eta_{\phi^2}=\gamma_{\phi^2}(y_\ast,\lambda_\ast)$.  As our main result,  we obtain at order $\epsilon^3$ for the inverse correlation length exponent and the anomalous dimensions
\begin{widetext}
\begin{align}
\frac{1}{\nu}=&2-\frac{(3+10 N+s) \epsilon }{6 (3+2 N)}-\frac{513-7587 N-666 N^2-5264 N^3-96 N^4+s(171+510 N+436 N^2+48 N^3)}{108(3+2 N)^3s} \epsilon ^2\label{eq:numin1}\\
	&+\frac{\epsilon^3}{3888 (3+2 N)^5 s^3}\Big(-227691 (3+s)+4 N (81 (2170 s-128871)+N (27 (-2238507+554816 s)\nonumber\\
	&+2 N (585 (2414 s-16143)+N (4 N (5233698+N (1383001+16 N (3832+54 N-27 s)-24986 s)-371936 s)\nonumber\\
	&-3 (8117973+761116 s)))))+288 (3+2 N) s^2 (2 N (81+N (1917+4 N (450+N (153+4 N))))\nonumber\\
	&-N (3+4 N) (99+4 N (21+N)) s+81 (3+s)) \zeta _3\Big)\,.\nonumber\\[5pt]
	\eta_\psi=&\frac{\epsilon}{2 (3+2 N)}+\frac{180+33 s+N (3-328 N+2 s)}{216 (3+2 N)^3}\epsilon^2+\Big(\frac{102519+237519 N+342 N^2-122020 N^3-11040 N^4}{7776 (3+2 N)^5}\label{eq:etapsi}\\
	&-\frac{68607+2099304 N+1629828 N^2+1505352 N^3+89536 N^4+1248 N^5}{7776 (3+2 N)^5s}-\frac{6 (1+N) \zeta _3}{(3+2 N)^4}\Big)\epsilon^3\,,\nonumber\\[5pt]
	\eta_\phi=&\frac{2 N\epsilon}{3+2 N}+\frac{27+594N+2916 N^2+88 N^3+(9-57 N+208 N^2) s}{36 (3+2 N)^3s}\epsilon^2+\Big(\frac{2943+47385 N}{1296 (3+2 N)^5}-\frac{24 N(1 +N)}{(3+2 N)^4} \zeta _3\\
	&\hspace{-0.5cm}+\frac{8829-24192 N-1603476 N^2-292200 N^3-300224 N^4-6112 N^5+s(118926 N^2+115564 N^3+17312 N^4)}{1296 (3+2 N)^5s}\Big)\epsilon^3\,.\nonumber
\end{align}
\end{widetext}
%

\paragraph{Metal-insulator transition in graphene.}

One of the motivations behind this study is the improvement of our understanding of metal-insulator transitions in graphene and related Dirac materials.
The quantum phase transition from the semi-metallic state to the sublattice symmetry broken insulating state with charge order -- the charge density wave (CDW) -- belongs to the 3D~Gross-Neveu(-Yukawa) universality class\cite{herbut2006,PhysRevB.89.205403} for fermion flavor number $N=2$, corresponding to an eight-component spinor $\psi$. These components reflect the presence of two sublattices of the underlying honeycomb lattice $A$ and $B$, two inequivalent Dirac points in the Brillouin zone at $K$ and $-K$ and two spin species $(\uparrow,\downarrow)$.
The order parameter field is a spin singlet corresponding to a staggered density state with alternating charge densities on the different sublattices.
Condensation to a non-vanishing vacuum expectation value $\langle\phi\rangle\neq0$ spontaneously breaks the Lagrangian's chiral symmetry and induces a finite gap in the energy dispersion.
The quantum critical behavior of this transition for $N=2$ has previously been accessed by different approaches,  most recently by higher-order perturbative and non-perturbative RG calculations and Majorana Quantum Monte Carlo simulations: For the purely fermionic Gross-Neveu model expanded in $D=2+\epsilon$ dimensions the RG functions are known to order $\epsilon^4$ and for $\epsilon=1$ yield an inverse correlation length exponent $1/\nu\approx 0.931$ after resummation\cite{Gracey:2016mio}, while the most sophisticated non-perturbative functional RG calculation gives a value of $1/\nu\approx 0.994(2)$\cite{PhysRevB.94.245102}. Novel lattice methods have managed to access the question avoiding the sign problem, but predicting a rather different value of  $1/\nu\approx 1.20(1)$\cite{PhysRevD.88.021701}.
We show a numerical evaluation of Eq.~\eqref{eq:numin1} for $N=2$ order $\epsilon^3$ in Tab.~\ref{tab:critexp2}.
We note that, e.g., for the inverse correlation length exponent, the prefactor of the order $\epsilon^3$ term is larger than the one for the order $\epsilon^2$ term.
To obtain an estimate for the critical exponents at $\epsilon=1$, we first directly evaluate the series as given in  Tab.~\ref{tab:critexp2} at $\epsilon=1$ which gives $\nu^{-1}(\epsilon=1)\approx 0.960, \eta_\psi(\epsilon=1)\approx 0.0404$ and $\eta_\phi(\epsilon=1)\approx 0.667$.
Further, we employ the Pad\'e approximant [2/1] to obtain for $\nu^{-1}_{[2/1]}\approx 1.048$ for $\epsilon=1$.
The corresponding Pad\'e approximants for the fermion and boson anomalous dimensions are $\eta_{\psi [2/1]}\approx 0.0740$ and $\eta_{\phi [2/1]}\approx 0.672$, respectively.
Due to the considerable size of the order $\epsilon^3$ contribution the values from the direct substitution and the given Pad\'e approximants are spread over a sizeable interval.
We interpret this as a measure for the theoretical uncertainties.

\paragraph{Spinless fermions on the honeycomb lattice.}

Another much studied case is the one of spinless fermions on the honeycomb lattice, which also undergo a metal-insulator transition for strong repulsive interactions. It corresponds to the universality class which is realized for $N=1$ in our model.
Recent sign-problem free Monte Carlo studies\cite{PhysRevD.88.021701,1367-2630-16-10-103008,1367-2630-17-8-085003,PhysRevD.88.021701,PhysRevB.93.155157} and functional RG approaches\cite{PhysRevB.89.205403} give improved estimates on $\nu^{-1}, \eta_\phi$ and $\eta_\psi$, cf. Tab~\ref{tab:critexp}.
Again, for our order $\epsilon^3$ estimates at $\epsilon=1$, we employ direct substitution of $\epsilon=1$ and obtain $\nu^{-1}(\epsilon=1)\approx 1.099, \eta_\psi(\epsilon=1)\approx 0.0773$ and $\eta_\phi(\epsilon=1)\approx 0.439$.
The corresponding Pad\'e approximants [2/1] for $\epsilon=1$ are $\nu^{-1}_{[2/1]}\approx 1.166, \eta_{\psi [2/1]}\approx 0.102$ and $\eta_{\phi [2/1]}\approx 0.463$.

\paragraph{Emergent SUSY in topological superconductors.}

For $N=1/4$, the field content of the GNY model presented here is compatible with supersymmetry (SUSY) and an emergent SUSY scenario at the boundary of a topological phase that has been discussed in\cite{Grover:2013rc}. In this case a superscaling relation $1/\nu=(D-\eta)/2$ with $\eta=\eta_\psi=\eta_\phi$ is satisfied\cite{Gies:2009az} at the quantum critical point to all orders\cite{Heilmann:2014iga}.
Estimates for the critical exponents have been calculated with SUSY functional RG methods\cite{Heilmann:2014iga} as well as with the conformal bootstrap\cite{Bashkirov:2013vya,Iliesiu:2015akf}, cf. Tab.~\ref{tab:critexp}.
Within our calculations, we find that the super-scaling relation is exactly satisfied order by order for $D=4-\epsilon$ up to three loops and for the Pad\'e approximant [2/1] which gives for $\nu^{-1}_{[2/1]}\approx 1.419$ and for $\eta_{[2/1]}\approx 0.162$. For the Pad\'e approximant [1/2], we observe a violation of the superscaling relation. We interpret this as a hint towards a superior behavior of the Pad\'e approximant [2/1] over [1/2] for this model and order of the expansion, and then adopt this approximant also for the neighboring values of $N$,  as listed in Tab.~\ref{tab:critexp}.

\paragraph{Disorder-driven transition in Weyl semimetals.}

\begin{table}[t!]
\caption{\label{tab:critexp2} Critical exponents for the chiral Ising universality class to order $\epsilon^3$ for three different choices of $N$.}
\begin{tabular*}{\linewidth}{@{\extracolsep{\fill} } l l l l}
\hline\hline
$N=2:$  & $\nu^{-1}$ &$\approx$ &$2-0.952 \epsilon +0.00723 \epsilon ^2-0.0949 \epsilon ^3$\\
 & $\eta_{\psi}$ &$\approx$ &$0.0714 \epsilon -0.00671 \epsilon ^2-0.0243 \epsilon ^3$\\
 & $\eta_{\phi}$ &$\approx$ &$0.571 \epsilon +0.124 \epsilon ^2-0.0278 \epsilon ^3$\\[5pt]
 \hline
 $N=1:$  & $\nu^{-1}$ &$\approx$ &$2-0.835 \epsilon -0.00571 \epsilon ^2-0.0603 \epsilon ^3$\\
 & $\eta_{\psi}$ &$\approx$ &$0.1 \epsilon +0.0102 \epsilon ^2-0.033 \epsilon ^3$\\
 & $\eta_{\phi}$ &$\approx$ &$0.4 \epsilon +0.102 \epsilon ^2-0.0632 \epsilon ^3$\\[5pt]
  \hline
  $N=1/4:$  & $\nu^{-1}$ &$\approx$ &$2-0.571\epsilon -0.0204 \epsilon ^2+0.024 \epsilon ^3$\\
 & $\eta_{\psi}$ &$\approx$ &$0.143 \epsilon +0.0408 \epsilon ^2-0.048 \epsilon ^3$\\
 & $\eta_{\phi}$ &$\approx$ &$0.143 \epsilon +0.0408 \epsilon ^2-0.048 \epsilon ^3$\\[5pt]
  \hline
 $N=0:$  & $\nu^{-1}$ &$\approx$ &$2-0.333 \epsilon -0.117 \epsilon ^2+0.125 \epsilon ^3$\\
 & $\eta_{\psi}$ &$\approx$ &$0.167 \epsilon +0.0478 \epsilon ^2-0.0469 \epsilon ^3$\\
 & $\eta_{\phi}$ &$\approx$ &$0.0185 \epsilon ^2+0.0187 \epsilon ^3$\\
\hline\hline
\end{tabular*}
\end{table}
%

\begin{table}[t!]
\caption{\label{tab:critexp} Chiral Ising universality in $D=3$: Inverse correlation length exponent $1/\nu$ and anomalous dimensions  $\eta_\phi$  and $\eta_\psi$ for bosons and fermions, respectively. In {\it this work}, we provide results within the $(4-\epsilon)$ expansion to order~$\epsilon^3$.}
\begin{tabular*}{\linewidth}{@{\extracolsep{\fill} } c c c c c}
\hline\hline
$N=2$  & & $1/\nu$ & $\eta_\phi$ & $\eta_\psi$ \\
\hline
{\it this work} (Pad\'e $[2/1]$) & & 1.048 & 0.672 & 0.0740\\
$(2+\epsilon)$, ($\epsilon^4$, Pad\'e)\cite{Gracey:2016mio}& & 0.931 & 0.745 & 0.082\\
functional RG\cite{PhysRevB.94.245102}  & & 0.994(2) & 0.7765 & 0.0276\\
Monte Carlo\cite{PhysRevD.88.021701} & & 1.20(1) & 0.62(1) & 0.38(1)\\
\hline
$N=1$  & & $1/\nu$ & $\eta_\phi$ & $\eta_\psi$ \\
\hline
{\it this work} (Pad\'e $[2/1]$)  & & 1.166 & 0.463 & 0.102\\
functional RG\cite{PhysRevB.94.245102} & & 1.075(4) & 0.5506 & 0.0645\\
Monte Carlo\cite{1367-2630-17-8-085003} & & 1.30 & 0.45(3) &  \\
\hline
$N=1/4$  & & $1/\nu$ & $\eta_\phi$ & $\eta_\psi$ \\
\hline
{\it this work} (Pad\'e $[2/1]$)      &  &  1.419  & 0.162 & 0.162\\
functional RG\cite{Heilmann:2014iga} & & 1.408 & 0.180 & 0.180\\
conformal bootstrap\cite{Iliesiu:2015akf}  & &  & 0.164 & 0.164 \\
\hline\hline
\end{tabular*}
\end{table}
%

Another application concerns the replica limit $N\to 0$ which is believed to describe the transition from a relativistic semi-metallic state to a diffusive metallic phase in a 3D Weyl semi-metal. At one-loop order this unusual limit also gives rise to a NGFP which is non-trivial in both couplings $(y^\ast,\lambda^\ast)_+=(\epsilon/3,\epsilon/36)$.
Beyond one-loop order, the limit $N\to 0$ suffers from the lack of multiplicative renormalizability of the model and contributions from evanescent operators\cite{PhysRevB.94.220201,Gracey:2016mio}.
A previous four-loop expansion of the purely fermionic Gross-Neveu model in $D=2+\epsilon$ has shown very large four loop coefficients in comparison with the three-loop terms\cite{Gracey:2016mio}.
Here, we circumvent this problem by considering the GNY model in the same limit which allows us to provide three-loop estimates for the critical exponents of this disorder-induced transition.
Our order $\epsilon^3$ estimates, with direct substitution of $\epsilon=1$ yields $\nu^{-1}(\epsilon=1)\approx 1.673$ and $\eta_\psi(\epsilon=1)\approx 0.167$.
The Pad\'e approximants are $\nu^{-1}_{[2/1]}\approx 1.610$ and $\eta_{\psi [2/1]}\approx 0.191$.

\paragraph{Conclusions.} 

We have studied the Gross-Neveu-Yukawa model at three-loop order in $D=4-\epsilon$ space-time dimensions and have extracted the solution of the stable non-Gau\ss ian fixed point and the corresponding critical exponents to order~$\epsilon^3$ for arbitrary number $N$ of fermion flavors.
The model is believed to govern the universal critical behavior in a number of  quantum phase transitions in interacting or disordered Dirac fermions that are of current interest, and we provided estimates for the correlation length exponent and anomalous dimensions based on Pad\'e approximants for our three-loop series. Our calculation fully reproduces the previous two-loop results \cite{Rosenstein:1993zf} at all $N$, and the three-loop Ising model results \cite{Kleinert:2001ax} at $N=0$. For a single component fermion, coresponding to $N=1/4$ in our notation, our values of the exponents agree to all orders of the calculation with the scaling relation dictated by the supersymmetry that emerges in this limit. Finally, while the comparisons of the values of the critical exponents from different methods, i.e. Monte Carlo simulations, conformal bootstrap, perturbative and non-perturbative RG approaches, and our epsilon-expansion agree to a reasonable degree, they seem not to have fully converged yet.

In the future, it will also be interesting to study closely related field theories describing other patterns of symmetry breaking in presence of fermions\cite{PhysRevB.80.075432}, e.g., the antiferromagnetic transition in the Hubbard model on  honeycomb lattice, to compare the results with the large-scale Monte Carlo simulations \cite{PhysRevX.3.031010,PhysRevB.91.165108,PhysRevX.6.011029}.

\begin{acknowledgments}
The authors are grateful to Holger Gies for discussions. M.M.S. is supported by DFG Grant No. SCHE 1855/1-1 and I. F. H. is supported by the NSERC of Canada.
\end{acknowledgments}

%


\begin{thebibliography}{55}%

\bibitem{wehling2014}
  T.~Wehling, A.~Black-Schaffer, and A.~Balatsky, Advances in Physics {\bf 63}, 1 (2014).
  
\bibitem{2013arXiv1306.2272V}
  O.~Vafek and A.~Vishwanath, Annual Review of Condensed Matter Physics Vol. 5: 83-112 (2014).
  
\bibitem{Liu864}
 Z.~K.~Liu, B.~Zhou, Y.~Zhang, Z.~J.~Wang, H.~M.~Weng, D.~Prabhakaran, S.-K.~Mo, Z.~X.~Shen, Z.~Fang, X.~Dai, Z.~Hussain, and Y.~L.~Chen, Science {\bf 343}, 864 (2014).
  
\bibitem{PhysRevLett.113.027603}
 S.~Borisenko, Q.~Gibson, D.~Evtushinsky, V.~Zabolotnyy, B.~B\"uchner, and R.~J.~Cava, Phys. Rev. Lett. {\bf 113}, 027603 (2014).
  
\bibitem{sorella1992}
  S.~Sorella and E.~Tosatti, Europhysics Letters {\bf 19}, 699 (1992).

\bibitem{Khveshchenko2001}
 D.~V.~Khveshchenko, Phys. Rev. Lett. {\bf 87}, 246802 (2001).

\bibitem{herbut2006}
  I.~F.~Herbut, Phys. Rev. Lett. {\bf 97}, 146401 (2006).

\bibitem{honerkamp2008}
 C.~Honerkamp, Phys. Rev. Lett. {\bf 100}, 146404 (2008).
 
\bibitem{herbut2009}
  I.~F.~Herbut, V. ~{Juri\ifmmode \check{c}\else \v{c}\fi{}i\ifmmode~\acute{c}\else \'{c}\fi{}}, and B. Roy, Phys. Rev. B {\bf 79}, 085116 (2009).

\bibitem{raghu2008}
  S.~Raghu, X.-L.~Qi, C.~Honerkamp, and S.-C.~Zhang, Phys. Rev. Lett. {\bf 100}, 156401 (2008).

\bibitem{PhysRevB.33.3263}
E.~Fradkin, Phys. Rev. B {\bf 33}, 3263 (1986).

\bibitem{PhysRevB.33.3257}
E.~Fradkin, Phys. Rev. B {\bf 33}, 3257 (1986).

\bibitem{PhysRevLett.107.196803}
P.~Goswami and S.~Chakravarty, Phys. Rev. Lett. {\bf 107}, 196803 (2011).

\bibitem{PhysRevLett.112.016402}
  K.~Kobayashi, T.~Ohtsuki, K.-I.~Imura, and I.~F.~Herbut, Phys. Rev. Lett. {\bf 112}, 016402 (2014).

\bibitem{PhysRevB.94.220201}
T.~Louvet, D.~Carpentier, and A.~A.~Fedorenko, Phys. Rev. B {\bf 94}, 220201 (2016).

\bibitem{PhysRevB.93.155113}
 S.~V.~Syzranov, P.~M.~Ostrovsky, V.~Gurarie, and L.~Radzihovsky, Phys. Rev. B {\bf 93}, 155113 (2016).

\bibitem{PhysRevD.10.3235}
D.~J.~Gross and A.~Neveu, Phys. Rev. D {\bf 10}, 3235 (1974).

\bibitem{doi:10.1093/ptep/ptw120}
L.~Fei, S.~Giombi, I.~R.~Klebanov, and G.~Tarnopolsky,
Progress of Theoretical and Experimental Physics 2016,
12C105 (2016).

\bibitem{HASENFRATZ199179}
  A.~Hasenfratz, P.~Hasenfratz, K.~Jansen, J.~Kuti, and Y.~Shen, Nuclear Physics B {\bf 365}, 79 (1991).
  
\bibitem{ZINNJUSTIN1991105}
 J.~Zinn-Justin, Nuclear Physics B {\bf 367}, 105 (1991).

\bibitem{Rosenstein:1993zf}
 B.~Rosenstein, H.-L.~Yu, and A.~Kovner, Phys. Lett. B {\bf 314}, 381 (1993).
 
\bibitem{Zerf:2016fti}
  N.~Zerf, C.-H.~Lin, and J.~Maciejko, Phys. Rev. B {\bf 94}, 205106 (2016).
  
\bibitem{GRACEY1990403}
J.~A.~Gracey, Nuclear Physics B {\bf 341}, 403 (1990).

\bibitem{Gracey:2016mio}
  J.~A.~Gracey, T.~Luthe, and Y.~Schroder, Phys. Rev. D {\bf 94}, 125028 (2016).
  
\bibitem{PhysRevB.94.245102}
 B.~Knorr, Phys. Rev. B {\bf 94}, 245102 (2016).
 
\bibitem{PhysRevB.89.205403}
  L.~Janssen and I.~F.~Herbut, Phys. Rev. B {\bf 89}, 205403 (2014).
  
\bibitem{KARKKAINEN1994781}
  L.~K\"arkk\"ainen, R.~Lacaze, P.~Lacock, and B.~Petersson, Nuclear Physics B {\bf 415}, 781 (1994).

\bibitem{PhysRevD.88.021701}
  S.~Chandrasekharan and A.~Li, Phys. Rev. D {\bf 88}, 021701 (2013).
  
\bibitem{1367-2630-16-10-103008}
  L.~Wang, P.~Corboz, and M.~Troyer, New Journal of Physics {\bf 16}, 103008 (2014).

\bibitem{1367-2630-17-8-085003}
  Z.-X.~Li, Y.-F.~Jiang, and H.~Yao, New Journal of Physics {\bf 17}, 085003 (2015).
  
\bibitem{PhysRevB.93.155157}
S.~Hesselmann and S.~Wessel, Phys. Rev. B {\bf 93}, 155157 (2016).

\bibitem{Poland:2016chs}
  D.~Poland and D.~Simmons-Duffin, Nature Phys. {\bf 12}, 535 (2016).
  
\bibitem{Bashkirov:2013vya}
  D.~Bashkirov, arXiv:1310.8255 [hep-th] (2013).

\bibitem{Iliesiu:2015akf}
 L.~Iliesiu, F.~Kos, D.~Poland, S.~S.~Pufu, D.~Simmons-Duffin, and R.~Yacoby, JHEP {\bf 04}, 074 (2016).
 
\bibitem{0305-4470-31-40-006}
  R.~Guida and J.~Zinn-Justin, Journal of Physics A: Mathematical and General {\bf 31}, 8103 (1998).

\bibitem{PhysRevE.65.066127}
M.~Campostrini, A.~Pelissetto, P.~Rossi, and E.~Vicari, Phys. Rev. E {\bf 65}, 066127 (2002).

\bibitem{PhysRevE.68.036125}
  Y.~Deng and H.~W.~J.~Bl\"ote, Phys. Rev. E {\bf 68}, 036125 (2003).

\bibitem{PhysRevB.82.174433}
M.~Hasenbusch, Phys. Rev. B {\bf 82}, 174433 (2010).

\bibitem{Kos2016}
 F.~Kos, D.~Poland, D.~Simmons-Duffin, and A.~Vichi, JHEP {\bf 08} 036 (2016).

\bibitem{NOGUEIRA1993279}
P. Nogueira, Journal of Computational Physics {\bf 105}, 279 (1993).

\bibitem{Harlander1998125}
 R.~Harlander, T.~Seidensticker, and M.~Steinhauser, Physics Letters B {\bf 426}, 125 (1998).

\bibitem{Seidensticker:1999bb}
  T.~Seidensticker, arXiv:hep-ph/9905298 [hep-ph]  (1999).
  
\bibitem{Vermaseren:2000nd}
  J.~A.~M.~Vermaseren, arXiv:math-ph/0010025 [math-ph] (2000).

\bibitem{Kuipers20131453}
  J.~Kuipers, T.~Ueda, J.~Vermaseren, and J.~Vollinga,
Computer Physics Communications {\bf 184}, 1453 (2013).

\bibitem{Larin:1991fz}
  S.~A.~Larin, F.~V.~Tkachov, and J.~A.~M.~Vermaseren, NIKHEF-H-91-18 (1991).
  
\bibitem{Chetyrkin:1997fm}
  K.~G.~Chetyrkin, M.~Misiak, and M.~Munz, Nucl. Phys. B {\bf 518}, 473 (1998).

\bibitem{STEINHAUSER2001335}
  M.~Steinhauser, Computer Physics Communications {\bf 134}, 335 (2001).

\bibitem{Kleinert:2001ax}
 H.~Kleinert and V.~Schulte-Frohlinde, {\it Critical properties of phi**4-theories} (2001).
 
\bibitem{PhysRevB.80.075432}
  I.~F.~Herbut, V. ~{Juri\ifmmode \check{c}\else \v{c}\fi{}i\ifmmode~\acute{c}\else \'{c}\fi{}}, and O.~Vafek, Phys. Rev. B {\bf 80}, 075432 (2009).
  
\bibitem{Grover:2013rc}
 T.~Grover, D.~N.~Sheng, and A.~Vishwanath, Science {\bf 344}, 280 (2014).
 
\bibitem{Gies:2009az}
 H.~Gies, F.~Synatschke, and A.~Wipf, Phys. Rev. D {\bf 80}, 101701 (2009).

\bibitem{Heilmann:2014iga}
  M.~Heilmann, T.~Hellwig, B.~Knorr, M.~Ansorg, and A.~Wipf, JHEP {\bf 02}, 109 (2015).

\bibitem{PhysRevX.3.031010}
  F.~F.~Assaad and I.~F.~Herbut, Phys. Rev. X {\bf 3}, 031010 (2013).

\bibitem{PhysRevB.91.165108}
 F.~Parisen Toldin, M.~Hohenadler, F.~F.~Assaad, and I.~F.~Herbut, Phys. Rev. B {\bf 91}, 165108 (2015).

\bibitem{PhysRevX.6.011029}
  Y.~Otsuka, S.~Yunoki, and S.~Sorella, Phys. Rev. X {\bf 6}, 011029 (2016).

\end{thebibliography}
\end{document}